\documentclass[fleqn,12pt]{wlscirep}
\usepackage[utf8]{inputenc}
\usepackage[T1]{fontenc}
\usepackage{color}
\DeclareUnicodeCharacter{00A0}{ }

\usepackage{subcaption}
\usepackage{mwe}
 \usepackage{gensymb}
 \usepackage{lineno}
 
\title{Jets from MRC\,0600$-$399 bent by magnetic fields in the cluster Abell 3376} 
\author[1,2,$\dagger$,*]{James O. Chibueze}
\author[3,$\ddagger$,*]{Haruka Sakemi}
\author[3,+,*]{Takumi Ohmura}
\author[4]{Mami Machida}
\author[5]{Hiroki Akamatsu}
\author[6]{Takuya Akahori}
\author[7]{Hiroyuki Nakanishi}
\author[8,9]{Viral Parekh}
\author[8]{Ruby van Rooyen}
\author[10,11]{Tsutomu T. Takeuchi}

\affil[1]{Centre for Space Research, Potchefstroom campus, North-West University,
Potchefstroom 2531, South Africa}
\affil[2]{Department of Physics and Astronomy, Faculty of Physical Sciences,  University of Nigeria, Carver Building, 1 University Road,
Nsukka 410001, Nigeria}
\affil[3]{Graduated School of Science, Kyushu University, 744 Motooka Nishi-ku, Fukuoka, Fukuoka 819-0395, Japan}
\affil[4]{Division of Science, National Astronomical Observatory of Japan, 2-21-1 Osawa, Mitaka, Tokyo 181-0015, Japan}
\affil[5]{SRON Netherlands Institute for Space Research, Sorbonnelaan 2, 3584 CA Utrecht, The Netherlands}
\affil[6]{Mizusawa VLBI Observatory, National Astronomical Observatory of Japan, 2-21-1 Osawa, Mitaka, Tokyo 181-0015, Japan}
\affil[7]{Graduate School of Science and Engineering, Kagoshima University, 1-21-35 Korimoto Kagoshima 890-0065, Japan}
\affil[8]{South African Radio Astronomy Observatory, The Park, Park Road, Pinelands, 2 Fir Street, Black River Park, Observatory, 7925, South Africa}
\affil[9]{Department of Physics and Electronics, Rhodes University, PO Box 94, Makhanda, 6140, South Africa}
\affil[10]{Division of Particle and Astrophysical Science, Nagoya University, Furo-cho, Chikusa-ku, Nagoya 464-8602, Japan}
\affil[11]{The Research Center for Statistical Machine
Learning, the Institute of Statistical Mathematics, 10-3 Midori-cho, Tachikawa, Tokyo 190-8562, Japan}

\affil[$\dagger$]{james.chibueze@nwu.ac.za}
\affil[$\ddagger$]{sakemi@phys.kyushu-u.ac.jp}
\affil[+]{ohmura@phys.kyushu-u.ac.jp}
\affil[*]{Corresponding authors. These authors contributed equally to this work.}

\begin{abstract}

{\bf Galaxy clusters are known to harbour magnetic fields. 
The nature of the intra-cluster magnetic fields remains an unresolved question.
Intra-cluster magnetic field can be observed at the density contact discontinuity formed by cool and dense plasma running into hot ambient plasma~[1,2],
and the discontinuity exists[3]
near the 2nd BCG MRC 0600-399[4] of a merging galaxy cluster Abell\,3376 ($z=0.0461$, hereafter as A3376).
Elongated X-ray image in the east-west direction with a comet-like structure reaches a Mpc-scale[5] (Fig 1.(a)).
Previous radio observations[6,7] detected the bent jets from MRC 0600-399, moving in same direction as the sub-cluster's motion against ram pressure.

Here we report \bf{a new radio} observation of a radio galaxy MRC 0600-399 which is 3.4 times and 11 times higher resolution and sensitivity than the previous results[6].
Contrary to typical jets[8,9], the MRC 0600-399 shows a 90$^{\degree}$ bend at the contact discontinuity and the collimated jets further extend over 100 kpc from the bend point.
Diffuse, elongated emission named “double-scythe” structures were detected for the first time.
The spectral index flattens downstream of the bend point, indicating cosmic-ray re-acceleration. High-resolution numerical simulations reveal that the ordered magnetic field along the discontinuity 
plays a significant role in the change in the jet direction. The morphology of the “double-scythe” bear remarkable similarities with the simulations, which strengthens our understanding of the interaction between relativistic electrons and intra-cluster magnetic field.
}
\end{abstract}

\begin{document}

\flushbottom
\maketitle
%
%
\thispagestyle{empty}


In hierarchical structure formation, growth of galaxy clusters results in accretion and merger of sub-halos. This leads to shock waves, which heat up the plasma, disturb its motion, and accelerate cosmic-rays[10].

Thus, galaxy clusters are an ideal celestial laboratory to study plasma physics. An example of which is wide-angle tail (WAT) radio galaxies[8,9].
Another example is the so-called cold front, a density contact discontinuity of the intra-cluster medium (ICM) [1,2] resulting from gas motion.
Such a motion naturally strips the gas by ram pressure [11] and damps or amplifies magnetic fields[12].
Since their discovery,  WAT sources and cold fronts have been the focus of many studies to understand their origins, interaction with the ICM, and the nature of the intra-cluster magnetic field[13-17].

MRC 0600-399 was observed at 1.28 GHz with MeerKAT[18,19] to investigate its morphology~(Fig. 1).
In the center of A3376, there are two prominent radio galaxies - MRC 0600-399 ($z=0.04559$) and galaxy B ($z=0.0480$)(~Fig. 1. (b))
MRC 0600-399 also shows two-sided bent jets, but the collimated structures continue to the east direction over $\sim 100$~kpc and $\sim 50$~kpc for the northern and southern jets, respectively, beyond the bend points. 
The bend points refer to the locations 
as indicated with red dashed circles in Fig. 1(b).
The radio fluxes of the jets (especially the northern jet) drastically decreases before the bend points. There are also diffuse faint structures in the opposite (west) direction to the bent jets. We will refer to these as the ``double-scythe''. At the southern boundary of the northern bent jet, some diffuse filaments are resolved for the first time. These filaments appear faint compared to the rest of the emission, but are detected at well above the noise levels and indicated real structure related to the northern jet. These features are very unusual for typical WAT sources. 
We confirmed that there is no other radio galaxy overlapping with MRC 0600-399 and causing the ``double-scythe'' structure, thus, we confirm its association with the jets of MRC 0600-399.

Spectral index $\alpha$ map  and 1-dimensional profiles of the spectral index  $\alpha$ and flux density ($F_\nu \propto \nu^\alpha$) along the bent jets are shown in Fig.2 (a).
The spectral index values decrease gradually in N1 and S1 starting from MRC 0600-399. The spectral index decrement across 49~kpc (7 open circles) is $\sim 0.76$ for N1, and that across 25~kpc (4 open circles) is $\sim 0.44$ for S1. 
The trend then drastically changes and these values become fairly constant in N2 and S2, suggesting re-acceleration of relativistic electrons. Finally, these values again decrease in N3 with the decrement of $\sim 0.69$ across 30~kpc (6 open circles). The flux density values show a similar trend with the spectral indices in each part. The flux densities of the radio emission of the jets gradually decreases, and in the bend regions, the flux densities clearly increase. 

The radial X-ray surface brightness profile across the northern bent jet (Fig. 3) b bottom insert) suggests a clear discontinuity, which we interpret as a cold front.
From earlier numerical simulations[20-22], thick magnetic-field layers around cold fronts can be formed. Indeed, some observational evidence of amplified magnetic field across cold fronts has been reported[2,23].
This unusual radio morphology is ascribed to (partly re-accelerated) relativistic electrons traveling along with the magnetic fields of jets and the ICM over 150 kpc.
All observational evidence and previous numerical simulations point toward an interaction between jets and intra-cluster magnetic field along the cold front. 

In order to understand the bent jet quantitatively, we performed three-dimensional magneto-hydrodynamic (MHD) simulations of the interaction between the jet and the intra-cluster magnetic field using the CANS+ code [24]. We adopted an arch-shape magnetic field to reproduce the magnetic layer of the cold fronts. A jet launched from MRC 0600-399 travels straight with supersonic speed and hits the magnetic arch.
The motion of the jet across the arch is suppressed due to the arch's magnetic tension. The flow escapes along with the magnetic arch particularly toward the east direction because the tension against the flow is weaker due to the field inclination with respect to the jet injection direction; 
see Fig. 3 (a)
at the elapsed time of 110\,Myr from the time of interaction of the jet with the magnetic arch. The escaped flow has a sharp ``double-scythe" shape because the Kelvin-Helmholtz instability is suppressed by the strong arch magnetic-fields. The strong fields also produce a backflow of the jet. The backflow collides with the incoming jet and reduce its momentum, resulting in turbulence. As a result, the jet width reaches ten times the initial size around the bend point.
For the MRC 0600-399 case, a magnetic field of order 10\,$\mu$G is required to reproduce the observed radio morphology.

The synchrotron radio image made with our simulation successfully reproduces major characteristic features of the northern jet 
(Fig. 3 b).
Firstly, at the bend point, the simulation shows a "double-scythe" shape. Particularly, the eastern emission, which is produced by the majority of the escaped flow, have a close resemblance to the MeerKAT image, while the emission at the western area is more diffuse because of the presence of less gas along this direction. Secondly, the location of radio emission relative to the X-ray profile is consistent with the observed one 
(see Fig. 3 b insert).
Thirdly, the simulation reproduces the profile of the radio flux across N1 to N2. As for the southern jet, there are a handful of consistencies such as the ``double-scythe'' structure and an enhancement of radio flux at the bend point, suggesting 
the interaction between the jet and ordered magnetic fields. Meanwhile, the simulation shows a more smoothed arc-like structure at the tip of the jet. This difference could be due to the projection effect for a different viewing angle of the jet and/or more complex structure of magnetic fields.

Our MHD simulation reproduced a 50 kpc long eastern scythe of the northern jet, while the observed emission extends up to 100 kpc from the bend point.
This difference suggests that there may be some aspects of the plasma which are not considered in our simulations such as thermal conduction, decoupling between cosmic-rays and thermal plasma. Here we propose a feasible possibility to explain the observed feature, namely magnetic reconnection.
Our simulations indicate that the current sheet is generated at the contact area between the jet and magnetic field layer (Fig. 3 c)
once the jet collides with the cold front. 
This suggests that the magnetic energy can be efficiently converted to the cosmic-ray energy at N2 via magnetic reconnection and in-situ particle re-acceleration should occur there (Fig. 3 d).
Actually, the decrement of the spectral index, i.e. aging of relativistic electrons, halts at N2 (Fig. 2 b). Then, the spectral index and radio flux decrease again at N3. It is likely that we are observing active transportation of the cosmic-ray particles which were generated and/or re-accelerated by magnetic reconnection through the ordered magnetic field in the cold front.
The schematic of the proposed scenario is given in  Fig. 4.

Finally, we discuss an alternative scenario. The most simple model would happen if MRC\,0600-399 departs from the bulk motion of the sub-cluster and sees their surrounding gas flowing to the east (slingshot merger[25,26]). Because of its motion, the jets from the MRC\,0600-399 feel ram pressure from west to east direction. Therefore, this slingshot could generate the observed bent morphology unlike the general perspective of the merger merger. Even in this case, to explain the observed enhancement in the radio flux, spectral index and, the  “double-scythe” structure,  an interaction with the magnetic layer is likely required. Upcoming polarization studies will shed light on the nature of the bent jets.

The important aspects of our findings is that the bent jet in cold fronts could be a great probe to investigate the intra-cluster magnetic field. Ongoing/upcoming wide-field observations of galaxy clusters will reveal the nature of the intra-cluster magnetic field via the interaction with the jets.

\section{Figure Legends}

Figure 1: \textbf{Multi-wavelength view of Abell\,3376 and MRC 0600-399.}\\
(a) Composite image of Abell\,3376 (red-ish color: MeerKAT 1.28 GHz, lite-blue: X-ray, RGB: DSS-gir). To handle high-dynamic range image, the intensity of radio galaxies was scaled by 1/10. 
The white square indicates the main target of this work. 
Insert) The cyan arcs indicate the location of the cold front[3]. The solid cyan is this work. The red sector represent where the surface brightness was extracted.
(b) MeerKAT image of the total intensity of MRC 0600-399 at the center frequency of 1.28 GHz. The beam size is shown in bottom-left
corner, 5.80 $\times$ 5.48 arcsec$^{2}$. The magenta cross point indicates the position of the 2nd brightest cluster galaxy in optical band associated with MRC 0600-399. MRC 0600-399 has  jets which are bent 90 degree to the east, and continue to the east direction while keeping their collimated shapes. The arrows show the ``double-scythe'' structures, while the red dashed circles show the bend points. The yellow diamond point shows the position of optical source associated with galaxy B which is a D-type elliptical member galaxy of A3376. Galaxy B has also a two-sided jet, but they bend gently and the southern jet has a plume-like structure at the tail.

Figure 2: \textbf{Radio properties derived from MeerKAT observation.}\\
(a) Spectral index map derived from the radio datasets at 909-1658 MHz. Pixels with intensities below three times of total intensity rms levels are blanked.  Black circles on MRC 0600-399 indicate the regions in which the spectral index and flux density values are calculated and shown in bottom plots. The ellipse at bottom-left corner is the image resolution of 9.50 $\times$ 8.50 arcsec$^{2}$. (b) Plots of spectral indices (blue) and flux densities (red) of the regions at the northern (top) and southern (bottom) jets. The horizontal axes show the region numbers. Each region 1 of the northern and southern jets is above and below the boundary line respectively, shown in the spectral index map with a blue-dashed line. The blue-solid lines are the results of linear fitting of spectral indices in N1 and S1.

Figure 3: \textbf{
Numerical simulations of the jet--intra-cluster magnetic field interaction.}\\
(a) The 3D volume rendering of the absolute velocity for our MHD simulations at $t = 225 {\rm Myr}$. Yellow lines show the variation of magnetic fields from initial fields, $\delta B = B(t= 225~{\rm Myr}) - B(t = 0)$.
The box size is $234~{\rm kpc} \times 198~{\rm kpc} \times 126~{\rm kpc}$. 
(b) The projection plot shows the integrated synchrotron emissivity along a line of sight $\hat{n} = (0.06,0.71,-0.64)$ at $t = 168~{\rm Myr}$. The cyan sector represent where the surface brightness was extracted.
(insert top) The white solid and dashed lines represent the X-ray surface brightness and the radio intensity profile from our simulation. 
(insert bottom) Radial surface brightness profile of XMM-Newton data in  0.5--2.0 keV band across the northern bent radio emission. The blue solid and orange dashed lines represent the best-fit model of X-ray surface brightness profile and a radial profile of MeerKAT 1.28 GHz emission
(c) The color map (sliced from 3D data at $z=0$ ) shows the z-component of current density, $J_{\rm z} = (\nabla \times B)_{\rm z}$ at $t = 225~{\rm Myr}$.
(d) Map of joule heat ($\eta~J^2$) indicating regions of magnetic reconnection, where $\eta$ donates anomalous resistivity.

Figure 4: \textbf{Schematic drawing of the proposed scenario. }\\
As a galaxy cluster moving within the hot plasma sweep the intra-cluster magnetic fields, the magnetic field compresses along the contact discontinuity, forming a magnetic layer. AGN jet ejected from the member galaxy of the cluster receives a ram pressure by proper motion. 
Because the central AGN is well-inside of the 2nd BCG, MRC 0600-399, the ram pressure will not directly work on the jets.
When the jet terminal region reaches the magnetic layer on the cold front, the jet flow diverges along with the magnetic layer, called the ``double-scythe'' structure. Because the magnetic field in the AGN jet reconnects the magnetic layer, non-thermal particles accelerated by the magnetic reconnection propagate along with the magnetic layer. These particles emit synchrotron radiation. The northern part of the jet is located on the rim of the cold front, while the southern part hits it on the plane.

\section*{Methods}


We assume that Hubble's constant $H_0=70\; \rm km~s^{-1}~Mpc^{-1}$, the density parameter for mass $\Omega_{\rm M}=0.27$ and for dark energy $\Omega_{\rm \Lambda}=0.73$, which gives 54.5~kpc/arcmin at $z = 0.046$. Unless otherwise stated, the errors correspond to  68~\% confidence for each parameter.

\subsection*{MeerKAT L-band observations and data reduction}
Abell\,3376 East was observed with 60 antennas of the MeerKAT array [18,19] on 1 June, 2019 (Project ID: SCI-20190418-JC-01) at L-band (856 MHz to 1712 MHz). The MeerKAT array, located in the Northern Karoo desert of South Africa, is made up of 64 13.5-m ``offset Gregorian'' parabolic dish antennas. 48 of the 64 antennas are located in the inner core (within 1\,km radius) providing the shortest baseline of 29\,m, while the other 16 antennas are spread outside the core up to a maximum baseline of 8\,km. Thus, MeerKAT is capable of recovering a wide range (5$''$ to 27$'$) of angular scales at the central frequency of 1283 MHz.

The primary flux and bandpass calibrator was fixed to J0408-6545 (the total intensity, $I= 17$\,Jy at 1283 MHz). J0616-3456 ($I= 3.1$\,Jy at 1283 MHz, and 5.7$^o$ from the phase tracking center of Abell\,3376 East) was used as the secondary gain calibrator. During the observations, we performed four 10-minute scans of the primary calibrator, we scanned the secondary calibrator for every two minutes after the scan of the target. The flux, bandpass, and gain calibrations were reliably done with these bright calibrators.

The data correlation was done with the SKARAB correlator[27] in 4k mode with 856 MHz bandwidth, 4096 channels of $\sim$209 kHz per channel. Then, we have reduced the data with the semi-automated MeerKAT data analysis pipelines - OXKAT\footnote{https://ascl.net/code/v/2627}.

OXKAT is a semi-automatic pipeline (developed by Ian Heywood) used for MeerKAT data reduction and employs a collection of publicly available radio interferometry data flagging, calibration and imaging software packages. In the flagging process, the known radio frequency interference (RFI) channels, 856~MHz to 880~MHz, 1658~MHz to 1800~MHz, and 1419.8~MHz to 1421.3~MHz, are unexceptionally flagged out. Then, other possible RFIs are flagged using the autoflagger tricolour for the calibrators and by using AOFlagger [28] for the target fields. OXKAT pipeline uses the customary tasks from the CASA[29] suite for cross-calibration.

To deconvolve and image the target data, the WSClean imager [30] with Briggs weighting and a robust parameter of $-$0.3 was used, with the multi-scale and wide-band deconvolution algorithms enabled to better allow imaging diffuse emission present in the our fields. Deconvolution was performed in ten sub-band images of each 107 MHz wide-band. WSClean generates the multi-frequency synthesis (MFS) map (full bandwidth map), in joined-channel deconvolution mode, has central frequency of 1283 MHz. OXKAT pipeline uses the customary tasks from the Cubical software[31] for self-calibration.

We achieved synthesized beam of $5''.8 \times 5''.5$ and the rms noise level is 4.2 $\mu$Jy\,beam$^{-1}$ in the MFS image. The center frequency of the sub-band images are 909, 1016, 1123, 1230, 1337, 1444, 1551, and 1658 MHz. To derive the spectral index map (Fig. 2 b), we smoothed the resolutions of the sub-band images to the resolution ($9''.5 \times 8''.5$) of 909\,MHz sub-band image using CASA "imsmooth” task.

\subsection*{\textit{XMM-Newton}\label{xray} observation and Surface brightness profile analysis}

The radio galaxy was observed with EPIC instruments on the XMM-Newton X-ray observatory at 2003  (OBSID: 0151900101). The data was reduced with the science analysis system (SAS) and extended source analysis software (ESAS) packages, following the processing described in[3]. The tools emchain and epchain were used to create the reduced EPIC-MOS and pn event files, respectively. Light curves were extracted with 100s bins and screened for background flares with mos-filter and pn-filter. The cleaned exposure time for two MOS and pn data are 18.5 ks and 18.5 ks, respectively. The events with PATTERN values greater than 12 (MOS)/4 (pn) and non-zero FLAG values were excluded. The point sources with flux $>4\times10^{-14}~\rm erg/s/cm^2$ were excluded from the image. The resultant 0.5--2.0 keV X-ray image was shown in Fig. 1(a).

To investigate correlation with the bent radio emission, we fit the X-ray surface brightness profile with \texttt{Proffit}[32]. To model the observed profile, we used an underlying broken power-law density model:
\begin{align}
n(r) =
\begin{cases}
{C}n_0 \bigg (\dfrac{r}{r_{\rm edge}} \bigg )^{-\alpha_1} \, ,\quad &r \leq r_{\rm edge}
\\[12pt]
n_0 \bigg (\dfrac{r}{r_{\rm edge}} \bigg )^{-\alpha_2} \, ,\quad &r > r_{\rm edge} \, .
\end{cases}
\end{align}
where $\alpha$, $r_{\rm edge}$, $n_{0}$ and $C$ represent the slope of the profile, the location of the density jump, the normalization constant and compression factor (density ratio), respectively. The subscripts 1, 2 denote inside and outside of the edge of the density jump. The density model profile is projected into the line of sight assuming spherical geometry within the extraction sector. 

We fit the surface brightness profile in the annular shown in Fig. 3 (b). The opening angles 60$^\circ$--120$^\circ$ centered on  (6:02:13.014,-39:56:56.58) were selected to cover the bent jet structure.   The extracted surface brightness profile was radially binned to obtain SNR > 7. The observed profile was well reproduced with above model with $\chi^2$ = 93.2 for 83 d.o.f ($\Delta \chi=1.1$). The observed and the best-fitted model profiles are shown in Fig. 3 (b) insert. We find a surface brightness break across the region where the jet bends almost 90$^\circ$. The resultant density jump ($C=1.6\pm0.1$) is comparable to the previous report on the tip of the sub-cluster ($C$=1.8$\pm0.2$[3]). Since the presence of the cold front was confirmed at the tip of the sub-cluster, it is natural to consider the observed density jump is an extension of the cold front.

Intriguingly, radio emission shows a peak just outside the cold front, where the X-ray surface brightness slightly deviate from the expected values. Although the significance is low, this could be an indication of the presence of a plasma depletion layer due to intra-cluster magnetic field[33]. However, once we include an additional cut off around $r=2.5'$, the slope just outside of the first discontinuity is getting shallower. Hence the significance of the dip is also lower. In both cases, the too large  statistical errors prevent us from drawing any firm conclusion.

\subsection*{Numerical Simulation}

To study the interaction between jet and a magnetic layer behind cold front[34], we have conducted three dimensional MHD simulations using CANS+ code [24].

The units of length, velocity, density, pressure, temperature, and time in our simulations are $r_0 = 2~{\rm kpc}, v_0 = 3.9 \times 10^{2}~{\rm km~s^{-1}}, \rho_0 = 4.0 \times 10^{-27}~{\rm g~cm^{-3}}, p_0 = 6.2\times10^{-12}~{\rm erg~cm^{-3}}, kT_0 = 1~{\rm keV}$, and $t_0 = 4.8~{\rm Myr}$ respectively. Here, we take a mean molecular weight of 0.62.
We use the Cartesian domain $(-78 r_0, 78 r_0)\times(0, 96 r_0)\times( -31.5 r_0, 31.5 r_0)$ and uniform grids, whose size is $\Delta x=\Delta y = \Delta z = 0.15~r_0$. Here, the $y$ direction is aligned with jet axis. We impose a zero-gradient boundaries on all the sides of the box.

We focus mainly on the effect of the jet bending and collimating process by jet-cold front interaction.
Thus, we set a simple system that the jet propagate into a homogeneous static ICM and hit the arch-like strong fields.
The modeling of realistic cluster environments, such as turbulence fields and the inertia of moving cluster, is a future task.
To mimic accumulated magnetic fields along the cold fronts, we set a magnetic arch as follow:
\begin{equation}
    B_{\theta}(r) = B_{\rm ICM} + B_1 \sin{\left\{(r-r_{\rm s})\pi/ w \right\}},~~~B_{\rm r} = B_{\rm z} = 0, 
\end{equation}
where $r=\sqrt{x^2+y^2}$, $B_{\rm ICM} = 0.7~\mu{\rm G}$ is the ICM fields, $B_1 = 18~\mu{\rm G}$ is the peak fields of the magnetic arch at $r_s = 110$ kpc, $w = 30$ kpc is the width of the arch, respectively. 
The averaged fields of the magnetic arch are about $10$ $\mu$G, which is consistent with previous report of the cold fronts.
ICM pressure profile is determined by the equilibrium condition:
\begin{equation}
    p(r) = p(r=0) - \frac{B_{\theta}^2(r)}{8\pi} - \int^r_0{\frac{B_{\theta}^2(r)}{4\pi r}} dr,
\end{equation}
where $p(r=0) = 5 p_0$.
The ICM temperature is 5 KeV inside the magnetic arch[35].
The lowest value of plasma-$\beta$, $\beta \equiv 8\pi p/B^2$, is 1.3 at the top of the magnetic arch.
To satisfy the equilibrium condition, the pressure in the magnetic arch is $\sim~50$ \% lower than 
that of the inner area bounded by the arch.
This is inconsistent with observations and theories, which indicate that the pressure do not change across the cold fronts.
Although thermal pressure gradient of our initial condition may act to inflate the jets easier than realistic cluster, it do not play a role in the jet bending and collimating process.

The jet is injected at $(x,y,z) = (30r_0, 0, 0)$, with a radius, $r_{\rm jet} = 3$ kpc. The density, pressure, and velocity of injected flow are $\rho_{\rm jet} = 0.01 \rho_0, p_{\rm jet} = 5 p_0$, and $v_{\rm y,jet} = 1.5 \sqrt{\gamma p_{\rm jet}/\rho_{\rm jet}}$, respectively. 
The jet is injected with purely toroidal magnetic field 
\begin{equation}
    B=B_{\rm jet}~\sin^{4}(2\pi r'/r_{\rm jet})
\end{equation}

where $B_{\rm jet} = 5.6~ \mu$G and $r' = \sqrt{(x-30 r_0)^2 + z^2 }$. Therefore, the jet thermal, kinetic and magnetic energy luminosities are about 2.0 $\times 10^{43}$ erg s$^{-1}$, 2.5 $\times 10^{43}$ erg s$^{-1}$, and 6.8 $\times 10^{43}$ erg s$^{-1}$, respectively.
Meanwhile, the total energy of jetted gas, $e_{\rm jet} = 0.5\rho_{\rm jet}v^2_{\rm jet} + (\gamma -1 )^{-1}p_{\rm jet} + B^2_{\rm jet}/8\pi$, is  1.0 $\times 10^{-10}$ erg.
Contrary to this, the maximum magnetic energy of magnetic is 1.3 $\times 10^{-11}$ erg.
Thus, the magnetic energy is weaker than the jet total energy (13\% of the total energy).

We use a passive tracer function, $f(x,y,z,t)$, which is injected with the jet to divide it from that of the ICM. The tracer function have initially zero value elsewhere.  To clarify the distribution of jet plasma after the interaction, the tracer function takes the values of 1.0 in the injected region after the jet interact with the magnetic arch, $t \sim 115~{\rm Myr}$.

We model the radio by integrating the emissivity along the line of sight. The synchrotron emissivity, dropping physical constants, are then give by [36]
\begin{equation}
    \epsilon_{\rm sy} = N B_{\perp}^{\frac{1}{2}(\alpha+1)},
\end{equation}
where $N, B_{\perp}, \alpha = 0.5$ is the number density of relativistic electrons and the magnetic field perpendicular to the projection of the sky, and the power-law synchrotron spectral index, respectively. 
We model the population of relativistic electrons assuming it correlates with the product of the jet tracer function f (as described above) and the pressure $p$ of the gas[37].

We also calculate X-ray surface brightness to investigate correlation between radio jets and a location of cold front. Although the 0.5 - 2.0 KeV X-ray emissivity is roughly proportional to $n_{\rm e}^2$, we set a uniform ICM in our simulation. Therefore, we factorize the density as $n_{\rm e} = n_{\rm e, background} \times (\rho/\rho_0)$[38]. Here, $n_{\rm e, background}$ is the background density profile, and we adopt the $\beta$-model:
\begin{align}
n_{\rm e, background}(r) =
\begin{cases}
{C} n_0 \left\{ 1 + \left(r/r_{\rm c} \right) \right\}^{-3\beta/2}  ,\quad &r \leq r_{\rm edge}
\\[12pt]
n_0 \left\{ 1 + \left(r/r_{\rm c} \right) \right\}^{-3\beta/2} ,\quad &r > r_{\rm edge} \, ,
\end{cases}
\end{align}
where $r=\sqrt{x^2+y^2+z^2},~C = 1.6, n_0 = 10^{-3}~{\rm cm^{-3}},~r_{\rm c} = 25~{\rm kpc}, r_{\rm edge} = 110~{\rm kpc}$, respectively.
$r_{\rm edge}$ is the location of cold front assuming that the edge of magnetic layer.

To indicate regions where magnetic reconnection occur, we estimate the amount of joule heat ($=\eta J^2$). Because fast reconnections are observed in localized regions where the electric resistivity becomes anomalously high[39], we use anomalous resistivity as follows:
\begin{align}
\eta = \begin{cases}
1 ,\quad & v_{\rm d} >  v_{\rm d,crit}
\\[12pt]
0 ,\quad & v_{\rm d} \leq  v_{\rm d,crit} \, ,
\end{cases}
\end{align}
Where $v_{\rm d} \equiv |J|/\rho$ is electron drift velocity and $v_{\rm d,crit} = 5v_{0}$ is the critical velocity. High resistive region appears inside the jet. Although our post calculation to obtain the resistivity assumes the anomalous resistivity, it becomes almost constant due to high magnetic field strength and velocity. Dissipation inside the jet works as Ohmic dissipation rather than magnetic reconnection.

\section*{Data availability}
The raw MeerKAT data used for paper can be accessed from SARAO Archive \url{https://archive.sarao.ac.za} (Project ID: SCI-20190418-JC-01). The calibrated MeerKAT data and images that support the findings of this study are available from the corresponding authors upon reasonable request.




\section*{Acknowledgements}

JOC acknowledges support from the Italian Ministry of Foreign Affairs and International Cooperation (MAECI Grant Number ZA18GR02) and the South African Department of Science and Technology’s National Research Foundation (DST-NRF Grant Number 113121) as part of the ISARP RADIOSKY2020 Joint Research Scheme. VP acknowledges the financial assistance from the South African Radio Astronomy Observatory (SARAO) and the South African Research Chairs Initiative of the Department of Science and Technology and National Research Foundation. This paper makes use of the MeerKAT data (Project ID: SCI-20190418-JC-01). The MeerKAT telescope is operated by the South African Radio Astronomy Observatory, which is a facility of the National Research Foundation, an agency of the Department of Science and Innovation (DSI). SRON is supported financially by NWO, the Netherlands Organization for Scientific Research. Numerical computations and analyses were partially carried out on Cray XC50 and analysis servers at Center for Computational Astrophysics, National Astronomical Observatory of Japan, respectively. The computation was carried out using the computer resource by Research Institute for Information Technology, Kyushu University. This work was supported by JSPS KAKENHI Grant Numbers HS: 20J13339, TO: 20J12591, MM: 19K03916, and TTT: 17H01110 and 19H05076. 
TTT has also been supported in part by the Sumitomo Foundation Fiscal 2018 Grant for Basic Science Research Projects (180923), and the Collaboration Funding of the Institute of Statistical Mathematics ``New Development of the Studies on Galaxy Evolution with a Method of Data Science''. SAOImageDS9 development was made possible by funding from the Chandra X-ray Science Center (CXC), the High Energy Astrophysics Science Archive Center (HEASARC) and the JWST Mission office at the Space Telescope Science Institute[40].
This research used of Astropy, a community-developed core Python package for Astronomy[41].

\section*{Author contributions statement}
 J.O.C. conducted the observations and data reduction. V.P. also participated in the MeerKAT data reduction, while H.S. analysed the results and digested their implementations.
 T.O. and M.M. constructed theory, model and, conducted the numerical simulations. 
 H.A. performed X-ray data analysis and the scientific discussions.
 T.A. contributed to writing the MeerKAT proposal and the scientific discussions. T.T.T., R.vR., and H. N. contributed to the scientific discussions. 
 All authors reviewed the manuscript. 

\section*{Additional information}

\textbf{Competing interests}: The authors declare no competing interests. 



\begin{figure}[ht]
\centering

\includegraphics[width=\linewidth]{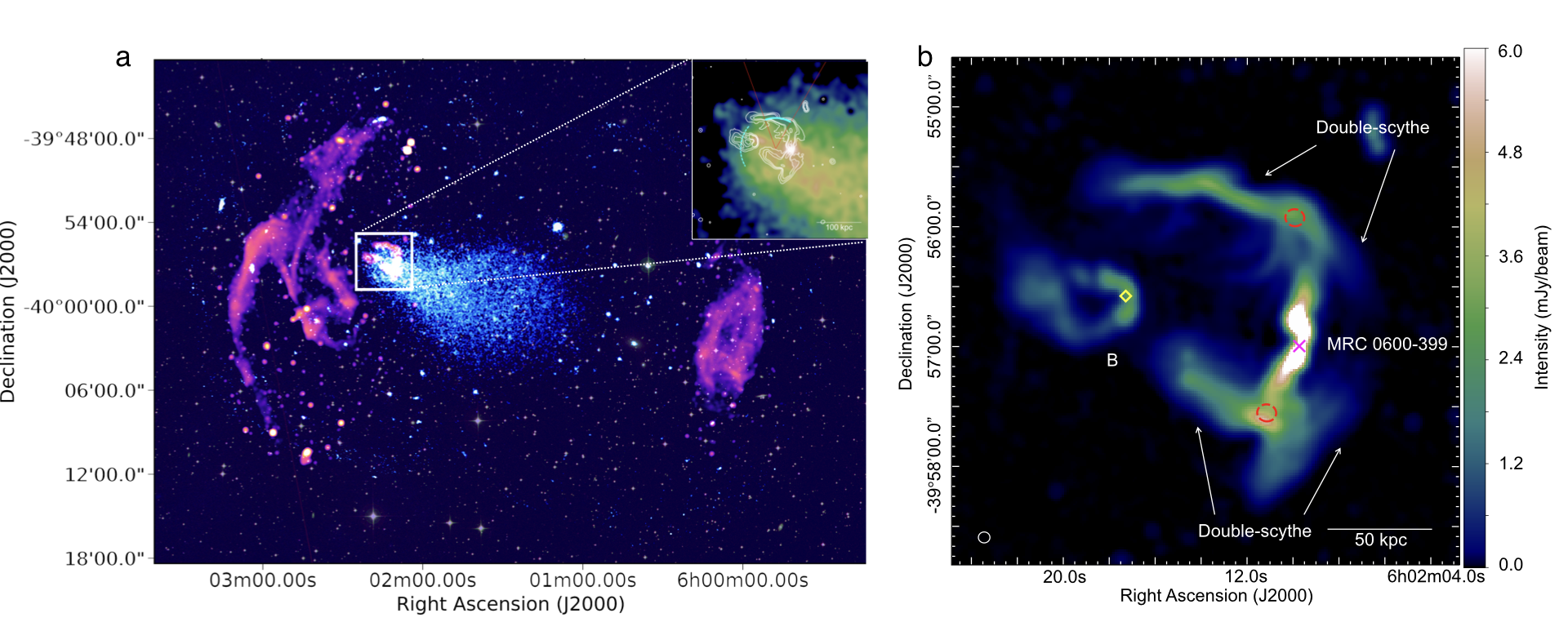}
\caption{
\label{fig:radio-fig} 
\textbf{Multi-wavelength view of Abell\,3376 and MRC 0600-399.}\\
(a) Composite image of Abell\,3376 (red-ish color: MeerKAT 1.28 GHz, lite-blue: X-ray, RGB: DSS-gir). To handle high-dynamic range image, the intensity of radio galaxies was scaled by 1/10. 
The white square indicates the main target of this work. 
Insert) The cyan arcs indicate the location of the cold front[3]. The solid cyan is this work. The red sector represent where the surface brightness was extracted.
(b) MeerKAT image of the total intensity of MRC 0600-399 at the center frequency of 1.28 GHz. The beam size is shown in bottom-left
corner, 5.80 $\times$ 5.48 arcsec$^{2}$. The magenta cross point indicates the position of the 2nd brightest cluster galaxy in optical band associated with MRC 0600-399. MRC 0600-399 has  jets which are bent 90 degree to the east, and continue to the east direction while keeping their collimated shapes. The arrows show the ``double-scythe'' structures, while the red dashed circles show the bend points. The yellow diamond point shows the position of optical source associated with galaxy B which is a D-type elliptical member galaxy of A3376. Galaxy B has also a two-sided jet, but they bend gently and the southern jet has a plume-like structure at the tail. 
}
\end{figure}

\begin{figure}[ht]
\centering
\includegraphics[width=\linewidth]{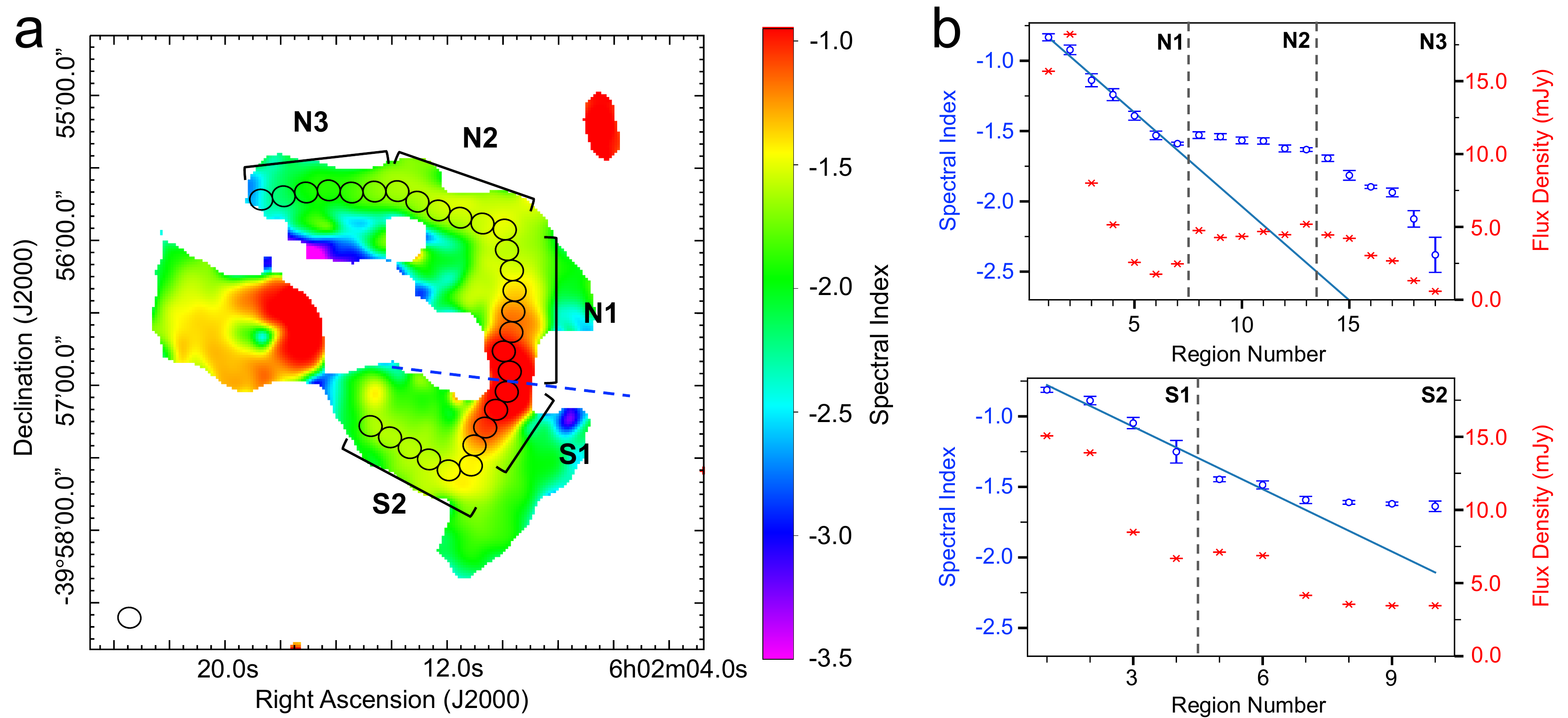}
\caption{
\label{fig:spec-fig}
\textbf{Radio properties derived from MeerKAT observation.}\\
(a) Spectral index map derived from the radio datasets at 909-1658 MHz. Pixels with intensities below three times of total intensity rms levels are blanked.  Black circles on MRC 0600-399 indicate the regions in which the spectral index and flux density values are calculated and shown in bottom plots. The ellipse at bottom-left corner is the image resolution of 9.50 $\times$ 8.50 arcsec$^{2}$. (b) Plots of spectral indices (blue) and flux densities (red) of the regions at the northern (top) and southern (bottom) jets. The horizontal axes show the region numbers. Each region 1 of the northern and southern jets is above and below the boundary line respectively, shown in the spectral index map with a blue-dashed line. The blue-solid lines are the results of linear fitting of spectral indices in N1 and S1.}
\end{figure}

\begin{figure}[ht]
\centering
\includegraphics[width=\linewidth]{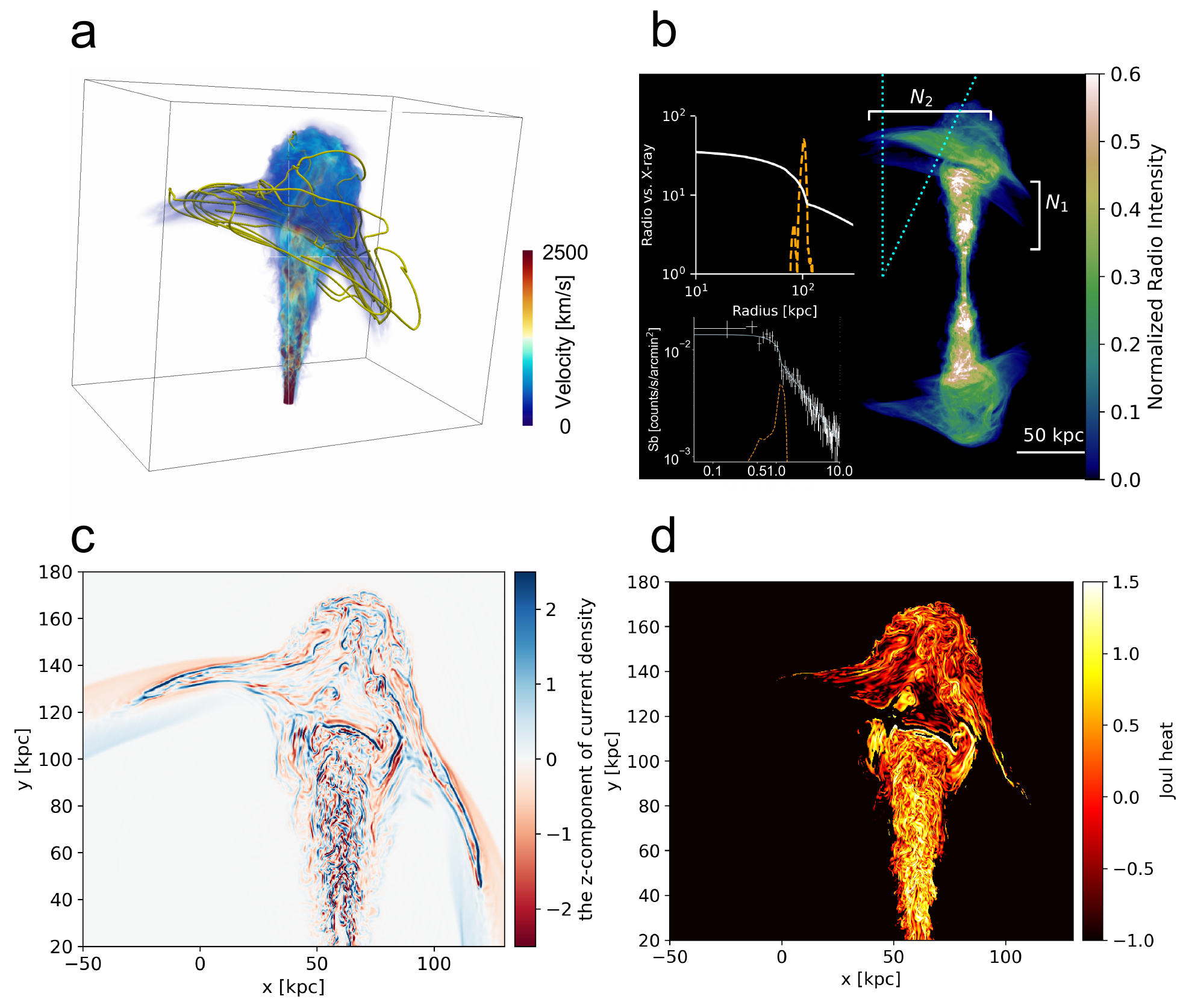}
\caption{
\label{fig:sim}
\textbf{
Numerical simulations of the jet--intra-cluster magnetic field interaction.}\\
(a) The 3D volume rendering of the absolute velocity for our MHD simulations at $t = 225 {\rm Myr}$. Yellow lines show the variation of magnetic fields from initial fields, $\delta B = B(t= 225~{\rm Myr}) - B(t = 0)$.
The box size is $234~{\rm kpc} \times 198~{\rm kpc} \times 126~{\rm kpc}$. 
(b) The projection plot shows the integrated synchrotron emissivity along a line of sight $\hat{n} = (0.06,0.71,-0.64)$ at $t = 168~{\rm Myr}$. The cyan sector represent where the surface brightness was extracted.
(insert top) The white solid and dashed lines represent the X-ray surface brightness and the radio intensity profile from our simulation. 
(insert bottom) Radial surface brightness profile of XMM-Newton data in  0.5--2.0 keV band across the northern bent radio emission. The blue solid and orange dashed lines represent the best-fit model of X-ray surface brightness profile and a radial profile of MeerKAT 1.28 GHz emission
(c) The color map (sliced from 3D data at $z=0$ ) shows the z-component of current density, $J_{\rm z} = (\nabla \times B)_{\rm z}$ at $t = 225~{\rm Myr}$.
(d) Map of joule heat ($\eta~J^2$) indicating regions of magnetic reconnection, where $\eta$ donates anomalous resistivity. 
}
\end{figure}

\begin{figure}[ht]
\centering
\includegraphics[width=0.6\linewidth]{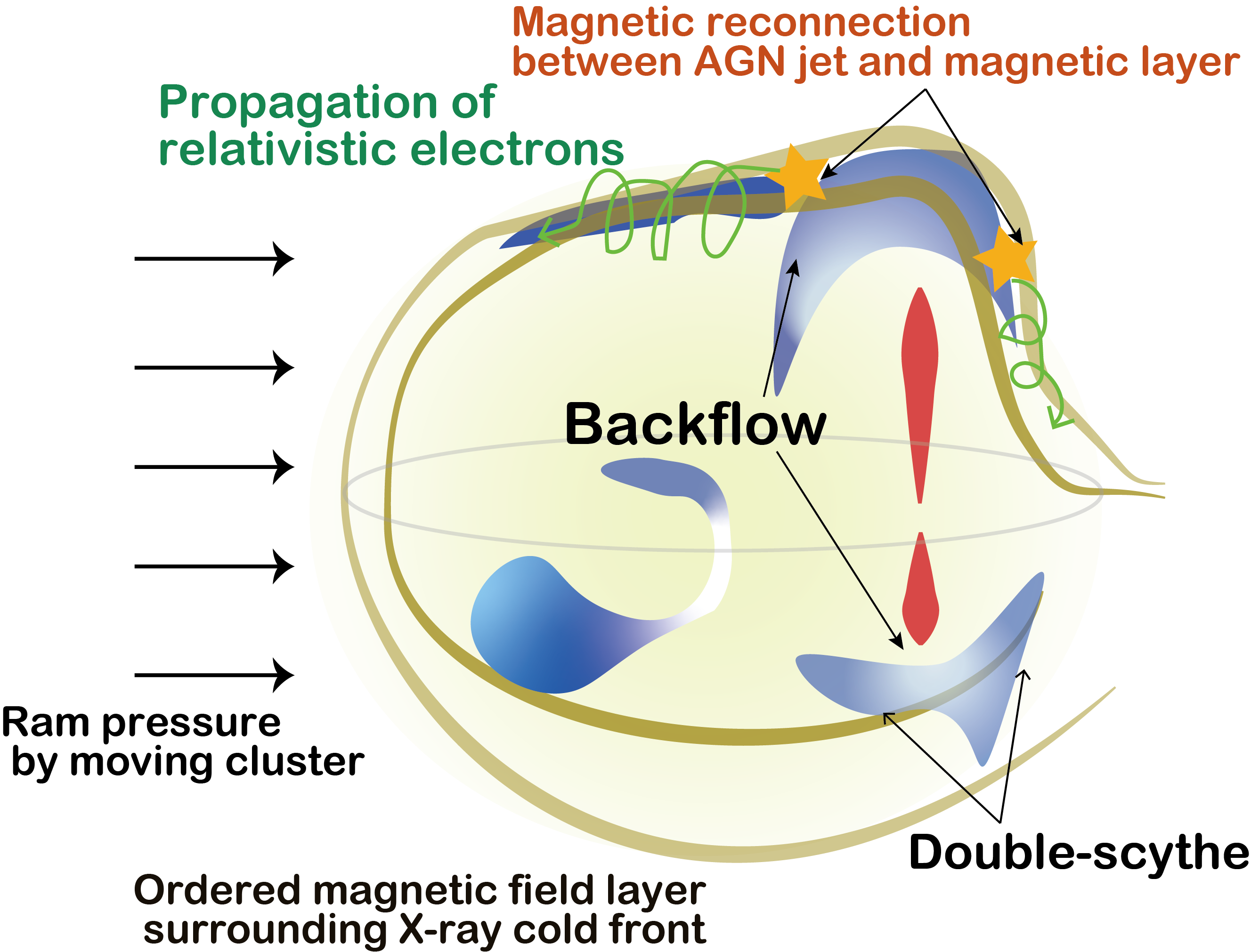}
\caption{
\label{fig:sch}
\textbf{Schematic drawing of the proposed scenario. }\\
As a galaxy cluster moving within the hot plasma sweep the intra-cluster magnetic fields, the magnetic field compresses along the contact discontinuity, forming a magnetic layer. AGN jet ejected from the member galaxy of the cluster receives a ram pressure by proper motion. 
Because the central AGN is well-inside of the 2nd BCG, MRC 0600-399, the ram pressure will not directly work on the jets.
When the jet terminal region reaches the magnetic layer on the cold front, the jet flow diverges along with the magnetic layer, called the ``double-scythe'' structure. Because the magnetic field in the AGN jet reconnects the magnetic layer, non-thermal particles accelerated by the magnetic reconnection propagate along with the magnetic layer. These particles emit synchrotron radiation. The northern part of the jet is located on the rim of the cold front, while the southern part hits it on the plane.}
\end{figure}

\end{document}